\documentclass[letterpaper]{emulateapj}
\usepackage{apjfonts,psfig,epsfig}

\lefthead{Smith, G.P., et al.}
\righthead{Evolution Since z=1 of the Morphology--density Relation}

\newcounter{fred}

\begin{document}

\def\zspec{\mathrel{z_{\rm spec}}}
\def\feso{\mathrel{f_{\rm E{+}S0}}}
\def\fso{\mathrel{f_{\rm S0}}}
\def\fe{\mathrel{f_{\rm E}}}
\def\ls{\mathrel{\hbox{\rlap{\hbox{\lower4pt\hbox{$\sim$}}}\hbox{$<$}}}}
\def\gs{\mathrel{\hbox{\rlap{\hbox{\lower4pt\hbox{$\sim$}}}\hbox{$>$}}}}

\title{Evolution Since z=1 of the Morphology--Density Relation for
  Galaxies }

\author{
Graham P.\ Smith,$\!$\altaffilmark{1,2}
Tommaso Treu,$\!$\altaffilmark{1,3}
Richard S.\ Ellis,$\!$\altaffilmark{1}\\
Sean M.\ Moran,$\!$\altaffilmark{1} and
Alan Dressler$\!$\altaffilmark{4}
}

\setcounter{footnote}{1}

\altaffiltext{1}{California Institute of Technology, Department of
                 Astronomy, Mail Code 105--24, Pasadena, CA 91125,
                 USA.}
\altaffiltext{2}{Email: gps@astro.caltech.edu}
\altaffiltext{3}{Hubble Fellow; Department of Physics
                 and Astronomy, University of California at Los
                 Angeles, Los Angeles, CA 90095, USA.}
\altaffiltext{4}{The Observatories of the Carnegie Institute of
                 Washington, 813 Santa Barbara Street, Pasadena, CA
                 91101, USA.}

\setcounter{footnote}{4}

\begin{abstract}
We measure the morphology--density relation of galaxies at $z{=}1$
across the full three orders of magnitude in projected galaxy density
available in low--redshift studies. Our study adopts techniques that
are comparable with those applied at lower redshifts, allowing a
direct investigation of how the morphological segregation of galaxies
has evolved over the last 8\,Gyr. Although the morphology--density
relation, as described by the fraction of early--type (E${+}$S0)
galaxies, was in place at $z{=}1$, its form differs from that observed
at both $z{=}0$ and $z{=}0.5$. In the highest density regions the
early--type fraction has increased steadily with time from
${\feso}{=}0.7{\pm}0.1$ at $z{=}1$ to ${\feso}{=}0.9{\pm}0.1$ at the
present epoch. However, in intermediate density regions corresponding
to groups and the accretion regions of rich clusters, significant
evolution appears to begin only after $z{=}$0.5. Finally, at the
lowest densities, no evolution is observed for the early type fraction
of field galaxies which remains constant at ${\feso}{=}0.4{\pm}0.1$ at
all epochs. We examine a simple picture consistent with these
observations where the early--type population at $z{=}1$ is comprised
largely of elliptical galaxies. Subsequent evolution in both
intermediate and dense regions is attributed to the transformation of
spirals into lenticulars. Further progress in verifying our hypothesis
may be achieved through distinguishing ellipticals and lenticulars at
these redshifts through resolved dynamical studies of representative
systems.
\end{abstract}


\section{Introduction}\label{intro}

In the local universe the fraction of galaxies with elliptical and
lenticular (i.e.\ early--type) morphologies is higher in clusters
of galaxies than in less dense environments (Hubble 1926; Oemler
1974; Melnick \& Sargent 1977; Dressler 1980). To first order,
this morphology--density relation appears to be a universal
characteristic of galaxy populations (e.g.\ Postman \& Geller
1984; Helsdon \& Ponman 2003). In quantitative terms,
morphological fractions correlate over three orders of magnitude
in projected galaxy density ($\Sigma$), thereby linking the
properties of cluster galaxies (${\Sigma}{\simeq}1000\,{\rm
Mpc^{-2}}$) with those of the field galaxy population
(${\Sigma}{\ls}10\,{\rm Mpc^{-2}}$) (Dressler 1980).

The morphological segregation of galaxies is a generic prediction
of cold dark matter simulations of large scale structure formation
(Frenk et al.\ 1985, 1988), and more recent semi--analytic galaxy
formation models (Kauffmann 1995; Baugh et al.\ 1996; Benson et
al.\ 2001; Diaferio et al.\ 2001).  In that context, the observed
morphology--density relation is interpreted as the combination of
two mechanisms. Firstly, the local density of galaxies and dark
matter is a proxy for the epoch of initial collapse of a given
structure; the most massive structures at any epoch represent the
earliest that collapsed. Secondly, interactions between galaxies,
dark matter and the intra--cluster medium (i.e.\ environmental
processes) are likely to transform in--falling field galaxies from
gas--rich spirals to gas--poor lenticular galaxies. The exact
balance between these two mechanisms (i.e.\ nature versus nurture)
and the detailed physics of the environmental processes have yet
to be identified unambiguously, and are the focus of much ongoing
research (e.g.\ Balogh et al.\ 2001; Kodama \& Smail 2001;  Treu et
al.\ 2003). 

An important element of investigating the physics of morphological
transformation is to trace the cosmic evolution of the
morphology--density relation over the full range of projected density
available locally. The timescales on which the relation evolves in
different density regimes will hold important clues to the physical
processes responsible.  To that end, Dressler et al.\ (1997) used
high--resolution imaging with the \emph{Hubble Space Telescope (HST)}
to measure the morphology--density relation in the core regions of a
sample of rich clusters at $z{\simeq}0.5$. Dressler et al. found that
the fraction of lenticular galaxies in clusters declined by a factor
of 2--3 between $z{=}0$ and $z{=}0.5$ and this evolution was
accompanied by a corresponding increase in the fraction of
star--forming spirals (see also Andreon 1998; Couch et al.\ 1998;
Fasano et al.\ 2000; Treu et al.\ 2003).

At higher redshifts, the distinction between elliptical and lenticular
morphologies becomes increasingly difficult to draw (Smail et al.\
1997; Fabricant et al.\ 2000).  Nevertheless, several authors have
measured the total early--type fraction ${\feso}$ in individual
clusters at $z{\simeq}1$ (e.g.\ van Dokkum et al.\ 2000, 2001; Lubin
et al.\ 2002).  These authors find ${\feso}{=}0.5$ in clusters at
$z{\simeq}1$, i.e.\ a smaller fraction than that found in the densest
environments at $z{=}0$.  However, as van~Dokkum \& Franx (2001)
caution, these estimates are preliminary because they are based on a
very small number of clusters.

In this paper we measure the morphology--density relation at
$z{=}1$ across the full three orders of magnitude in galaxy
density spanned in local samples.  We compare our results with
those obtained at lower and intermediate redshifts (Dressler 1980;
Dressler et al.\ 1997; Treu et al.\ 2003) and thus chart, for the
first time, the form of the morphology--density relation over a
cosmologically significant time interval (${\sim}8\,{\rm Gyr}$).

A plan of the paper follows. In \S\ref{data} we develop a strategy
for measuring the morphology--density relation at $z{=}1$ and
summarize the data used for this purpose. Then in \S\ref{analysis}
we describe the analysis, focusing separately on high-- and
low--density environments.  The main results, the
morphology--density relation at $z{=}1$ and its evolution to the
present--day are presented in \S\ref{results}. In \S\ref{discuss}
we discuss a possible interpretation, including how it relates to
previous measurements of $\feso$ in high--redshift clusters. We
summarize our conclusions in \S\ref{conc}.  We parameterize the
Hubble expansion as $h{=}H_0/100{\rm km s^{-1}Mpc^{-1}}{=}0.65$,
and adopt the currently favored values of $\Omega_{\rm M}{=}0.3$
and $\Omega_\Lambda{=}0.7$ when our analysis requires us to make
distance estimates.  In this cosmology $1''{\equiv}8.63{\rm kpc}$
physical size at $z{=}1$. Unless otherwise stated, all error bars
are stated at 1--$\sigma$ significance.  All magnitudes are quoted
in the Vega system.

\section{Data}\label{data}

\subsection{Strategy}\label{strategy}

The primary aims of this paper are to measure the morphology--density
relation at $z{\simeq}1$ and to identify broad evolutionary trends by
comparing our measurements with those at $z{\simeq}0$ (Dressler 1980)
and $z{\simeq}0.5$ (Dressler et al.\ 1997; Treu et al.\ 2003). To
facilitate this comparison, we adopt the same analysis methods used in
the lower redshift studies, and provide two measurements for carefully
selected galaxy populations at $z{=}1$ which form the basis of our
analysis. The projected number density (${\Sigma}$) of galaxies down
to $M_V{\le}M_V^\star+1$ allows us to measure the projected density,
${\Sigma}{\equiv}10/A$, where $A$ is the solid angle within which the
ten nearest neighbors are found (see \S\ref{analysis} for more
details). We also morphologically classify the galaxies in the various
$z{=}1$ samples. Both ${\Sigma}$ and morphologies need to be derived
in a homogeneous fashion across the full range in projected density:
$1{<}{\Sigma}{<}1000\,{\rm Mpc^{-2}}$.

Dressler et al.\ (1997) used \emph{HST} observations of 10
optically--selected clusters to measure the morphology--density
relation for cluster galaxies at $z{\simeq}0.5$, i.e.\
$50{\ls}{\Sigma}{\ls}1000\,{\rm Mpc^{-2}}$.  Treu et al.'s (2003)
wide--field (out to a projected cluster--centric radius of 5\,Mpc)
study of Cl\,0024 extends Dressler et al.'s results out to field
environments ${\Sigma}{\simeq}1\,{\rm Mpc^{-2}}$ for one cluster.  To
extend this body of work to $z{=}1$ we sought \emph{HST} imaging of a
similar sized sample of clusters at $z{\simeq}1$.  A search of the
\emph{HST} archive for WFPC2 observations of clusters at
$0.75{\le}z{\le}1.25$ through the F814W filter (i.e.\ a reasonable
match to rest--frame $V$--band) yielded a sample of six clusters for
which thirteen individual WFPC2 pointings are available
(Table~\ref{obs}).

To measure the morphological fractions at ${\Sigma}{\simeq}1\,{\rm
Mpc^{-2}}$, we complement these cluster data with a sample of
field galaxies. Prior to large--scale redshift surveys of galaxies
at $z{=}1$ in regions where \emph{HST} data is available (e.g.\
Davis et al.\ 2002; Le~F\`evre et al.\ 2003), we necessarily rely
on photometric redshift estimates.  We therefore selected a field
for which a deep photometric dataset with broad wavelength
coverage and \emph{HST} imaging through the F814W filter is
available.  The mosaiced \emph{HST} field containing the rich
cluster Cl\,0024 ($z{=}0.395$) is well--matched to this purpose as
the bulk of the faint population viewed is not associated with the
foreground cluster. Ground--based $BVRIJK$--photometry, plus F814W
\emph{HST}/WFPC2 imaging are available (Kneib et al.\ 2003) and
the projected physical extent is ${\sim}170{\rm Mpc^2}$ at
$z{=}1$, corresponding to a volume of $5{\times}10^5{\rm Mpc^3}$
when integrated over a redshift interval $0.75{\le}z{\le}1.25$.
Extensive spectroscopic studies of this field (Czoske et al.\
2001; Treu et al.\ 2003; Moran et al.\ 2004, in prep.) provide
several hundred spectroscopic redshifts which are useful in
calibrating photometric redshift estimates based on the
$BVRIJK$--band photometry (see \S\ref{losig}).

In comparing morphological fractions at different redshifts, in
addition to $k$--corrections and adopting a fixed luminosity limit of
$M_V^\star+1$ , the question of luminosity evolution in the population
needs to be considered. Interpolating between the redshift--dependent
$B$-- and $R$--band luminosity functions we estimate that evolution of
$M_V^\star$ between $z{=}1$ and $z{=}0$ is 1\,mag (Brown et al.\ 2001;
Chen et al.\ 2003; Norberg et al.\ 2002; Poli et al.\ 2003). Although
there is some uncertainty in this estimate, we conclude it is better
to apply this adjustment rather than to ignore the effect altogether.
We therefore subtract 1\,mag of evolution from $M_V^\star$ at $z{=}0$
(Brown et al.\ 2001), to define a luminosity limit 1\,mag fainter
than $M_V^\star$ at $z{=}1$, i.e.\ $M_V{\le}-21.2$.  

\begin{table}
\caption{Summary of \emph{HST} Data}
\label{obs}
\begin{center}
\begin{tabular}{lccccc}
\hline \noalign{\smallskip}
                   & {Redshift} & {$T_{\rm exp}({\rm ks})$} & {Pointings} & PID & {Reference} \cr
\noalign{\smallskip} \hline \noalign{\smallskip}
RCS\,0224${-}$0002 & 0.77 & {\hfil13.2} & 1 & 9135 & $a$ \cr
RXJ\,0848${+}$4453 & 1.27 & {\hfil28.0} & 1 & 6812 & $b$ \cr
MS\,1054${-}$0321 & 0.83 & {\hfil~~6.5} & 6 & 7372 & $c$ \cr
MS\,1137${+}$6624 & 0.78 & {\hfil15.0} & 1 & 5987  & \cr
Cl\,1325${+}$3009 & 0.76 & {\hfil19.0} & 1 & 6581  & $d$ \cr
Cl\,1604${+}$4304 & 0.90 & {\hfil19.0} & 3 & 8560 & $e$ \cr
\noalign{\smallskip} \hline \noalign{\smallskip}
Cl\,0024${+}$1654 & 0.395 & {\hfil~~4.4} & 38 & 8559 & $f$ \cr
Cl\,0024${+}$1654 & 0.395 & {\hfil18.0} & 1 & 5453 & \cr
\noalign{\smallskip} \hline \noalign{\smallskip}
\end{tabular}\\
\end{center}
{\footnotesize
$^a$~Gladders et al.\ (2002)\\
$^b$~van~Dokkum et al.\ (2001)\\
$^c$~van~Dokkum et al.\ (2000)\\
$^d$~Lubin, Oke \& Postman (2002)\\
$^e$~Postman, Lubin \& Oke (2001) -- these data include Cl\,1604${+}$4321.\\
$^f$~Treu et al.\ (2003) -- these data are used to characterize the
  $z{\simeq}1$ galaxy population, and not the galaxies that inhabit
  the foreground cluster at $z{=}0.4$.\\
}
\end{table}

\subsection{Space--based Observations}\label{space}

A wide--field sparse--sampled \emph{HST}/WFPC2\footnote{This paper
is based on observations with the NASA/ESA \emph{Hubble Space
Telescope}, obtained at the Space Telescope Science Institute
(STScI), which is operated by the Association of Universities for
Research in Astronomy, Inc., under NASA contract NAS5--26555.}
mosaic of Cl\,0024 ($z{=}0.395$) was acquired during Cycle~8
(PI:~R.S.\ Ellis, GO:8559), comprising 38 independent pointings
observed through the F814W filter for two orbits each.  Treu et
al.\ (2003) describe the reduction of these data; here we
summarize key details of the reduced data: the pixel--scale is
$0\farcs05$ after drizzling; the estimated $80\%$ completeness
limit is $I_{814}{\simeq}25$; the total combined field of view of
the 39 pointings (including the cluster center -- e.g.\ Smail et
al.\ 1997) is $0.05\,{\rm deg}^2$, excluding the PC chip from
each pointing.  The primary motivation of these observations was a
panoramic study of the rich cluster Cl\,0024 (Treu et al.\ 2003;
Kneib et al.\ 2003).  However, as discussed, these data provide
morphological information on a large sample of field galaxies at
$z{\simeq}1$ (\S\ref{morph}). The limiting magnitude of these data
corresponds to $M_V{\simeq}-20$ at $z{=}1$, i.e.\ sufficiently
deep to provide early/late--type morphological classification in a
manner consistent with that of earlier work (see \S\ref{morph} for
more details of the classification process, including estimation
of uncertainties).

The high--redshift cluster data (Table~\ref{obs}) were reduced using
the {\sc wfixup, wmosaic, imalign, imcombine} and {\sc cosmicrays}
tasks in {\sc iraf}\footnote{{\sc iraf} is distributed by the National
Optical Astronomy Observatories, which are operated by the Association
of Universities for Research in Astronomy, Inc., under cooperative
agreement with the National Science Foundation.}.  The reduced frames
have a pixel--scale of $0\farcs1$ and the mean FWHM of stellar
profiles is $0\farcs17$.  As this pixel scale is twice that of the
field--galaxy data described above, we block--averaged the field data
for the purpose of morphological classification.  Although this
results in a slight under--sampling of the WFPC2
point--spread--function, the larger pixels assist in the
identification of faint morphological features. Although these cluster
data are deeper than the corresponding field images (see
Table~\ref{obs}), both are sufficiently deep for the morphological
classification exercise (\S\ref{morph}).

\subsection{Ground--based Observations}\label{ground}

\begin{table}
\caption{Summary of Ground--based Data}
\label{tab-ground}
\begin{center}
\begin{tabular}{llcc}
\hline\noalign{\smallskip}
 {Filter} & Telescope/Instrument & 3--$\sigma$ Limit & FWHM($''$)\cr
\noalign{\smallskip}
\hline
\noalign{\smallskip}
$B$ & CFH12k & 27.8 & $0.92{\pm}0.06$ \cr
$V$ & CFH12k & 26.9 & $0.79{\pm}0.12$ \cr
$R$ & CFH12k & 26.6 & $0.88{\pm}0.09$ \cr
$I$ & CFH12k & 25.9 & $0.77{\pm}0.09$ \cr
$J$ & Hale/WIRC & 22.0 & $0.95{\pm}0.11$ \cr
$K$ & Hale/WIRC & 20.4 & $0.93{\pm}0.10$ \cr
\noalign{\smallskip}
\hline
\noalign{\smallskip}
\end{tabular}\\
\end{center}
\end{table}

\begin{figure}
\centerline{ \psfig{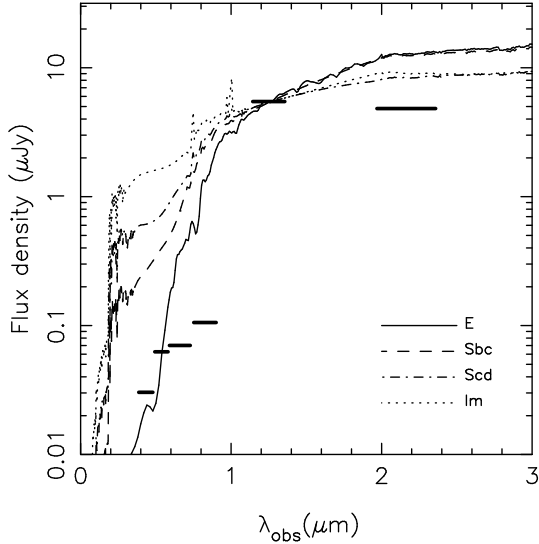} }

\caption{\small Observed detection limits for ground--based imaging of
the Cl\,0024 field, (3--${\sigma}$ significance, except for the
$J$--band, shown at 5--${\sigma}$). Coleman, Wu \& Weedman spectral
templates have been redshifted to $z{=}1$ and normalized to the
$V$--band luminosity of a galaxy at $z{=}1$ with
$M_V=M_V^\star+1=-21.2$ (see \S\ref{strategy} for details).  Our
ground--based data is shown to be sufficiently deep across the broad
wavelength range to achieve accurate photometric redshifts for all
spectral types (see \S\ref{ground} for further details).
\label{fig-detlim} }
\end{figure}

Panoramic optical data of Cl\,0024 were acquired with the 3.6--m
Canada France Hawaii Telescope\footnote{The Canada--France--Hawaii
Telescope (CFHT) is operated by the National Research Council of
Canada, l'Institut National des Science de l'Univers of the Centre
National de la Recherche Scientifique of France and the University
of Hawaii.} using the CFH12k camera (Cuillandre et al.\ 2000)
through the $BVRI$ filters.  These data are described by Czoske
(2002) and Treu et al.\ (2003).  The sensitivity limit and image
quality achieved in each passband is given in
Table~\ref{tab-ground}. The optical data are complemented by
wide--field $J$-- and $K_S$--band (hereafter $K$--band) imaging
obtained with the WIRC camera (Wilson et al.\ 2002) at the
Hale $200''$ telescope\footnote{The Hale Telescope at Palomar
Observatory is owned and operated by the California Institute of
Technology.} on 2002, October 29--30.  These near--infrared (NIR)
observations comprise a $3{\times}3$ mosaic of WIRC pointings,
providing a contiguous observed area of ${\sim}26'{\times}26'$
centered on the cluster.  Further details of these observations
and the data reduction are described by Kneib et al.\ (2003).
Here, we note that independent checks on the absolute photometric
calibration using unsaturated sources in the 2MASS point--source
and extended--source catalogs\footnote{This paper makes use of data
products from the Two Micron All Sky Survey (2MASS), which is a
joint project of the University of Massachusetts and the Infrared
Processing and Analysis Center/California of Technology, funded by
the National Aeronautics and Space Administration and the National
Science Foundation.}, together with examination of the sources
that fall in the overlap regions between the nine pointings
confirm that the absolute and relative calibration of both the
$J$-- and $K$--band data are accurate to $10\%$.  We incorporate
these uncertainties into the spectral template fitting described
in \S\ref{analysis}.  All of the ground--based data were
registered onto Czoske et al.'s (2001) astrometric grid, which is
accurate to ${\ls}0.2''$.

An important question is whether the depth of this multi--passband data
is adequate for reliable photometric redshift studies at $z{=}$1
described in \S\ref{losig}. We compare the depth of the ground--based
data as a function of wavelength to spectral templates derived from
observations of local galaxies (Coleman, Wu \& Weedman 1980 --
CWW). We redshifted the CWW templates to $z{=}1$ and normalized them
to $M_V=-21.2$ (see \S\ref{strategy}), and compared them with
detection limits listed in Table~\ref{tab-ground} (note that the
$J$--band detection limit is shown at 5--${\sigma}$ significance
because this is the detection filter adopted in \S\ref{losig}).
Fig.~\ref{fig-detlim} confirms that the ground--based data are
sufficiently deep to provide strong signal--to--noise detections
across the full wavelength range from $B$-- to $K$--bands for all but
the reddest spectral types.  The slight short--fall in sensitivity in
the bluest filters is not a significant concern because we have
ignored spectral evolution when constructing
Fig.~\ref{fig-detlim}. Indeed, only 3\% of the galaxies at
$z{\simeq}1$ in the final photometric redshift catalog are undetected
in the $B$--band.

\section{Analysis}\label{analysis}

In this section we describe how we construct samples of cluster
and field galaxies at $z{=}$1 and measure the projected density,
${\Sigma}$, at the location of each galaxy (\S\ref{env}). In
\S\ref{morph}, we describe the morphological classification.

\subsection{Measuring the Local Galaxy Density}\label{env}

\subsubsection{High Density Environments}\label{hisig}

We begin with the high density environments, using the pointed
WPFC2 observations of high redshift clusters (Table~\ref{obs}). We
analyzed each WFPC2 frame with SExtractor (Bertin \& Arnouts
1996), and adopted {\sc mag\_auto} as an estimate of the total
$I_{814}$--band magnitude of each source.  Assuming all the
detected sources are at the cluster redshifts (we discuss
corrections for contamination by field galaxies below), the total
$I_{814}$--band magnitudes were converted into the rest--frame
$V$--band using correction terms derived from synthetic spectral
templates for a representative range of stellar population ages
(2--8\,Gyr -- see Treu et al.\ (2001, 2003) for more details).  We
estimate that this step introduces an uncertainty of ${\ls}0.1{\rm
mag}$ in the estimated $M_V$ luminosity.  
We then select all galaxies with $M_V{\le}-21.2$ (i.e.\ the limit
defined in \S\ref{strategy}).

The projected number density was calculated for each of the 957
galaxies in the resulting catalog following the precepts introduced by
Dressler (1980). For each galaxy we counted the ten nearest neighbors
and divided by the rectangular area enclosed.  The median value of
$\Sigma$ computed in this manner is ${\Sigma}{\simeq}400\,{\rm
Mpc^{-2}}$; ${\sim}80\%$ of the galaxies have ${\Sigma}{\gs}200\,{\rm
Mpc^{-2}}$. Contamination arising from the projection of field
galaxies at lower and higher redshifts along the line--of--sight was
corrected using Postman et al.'s (1998) $I$--band number
counts. Given the broad bins in ${\Sigma}$ required to achieve
reasonable signal--to--noise (Fig.~\ref{fig-tsig}), uncertainties
arising from this correction do not significantly affect the final
cluster--based results.

\subsubsection{Low Density Environments}\label{losig}

\begin{figure*}
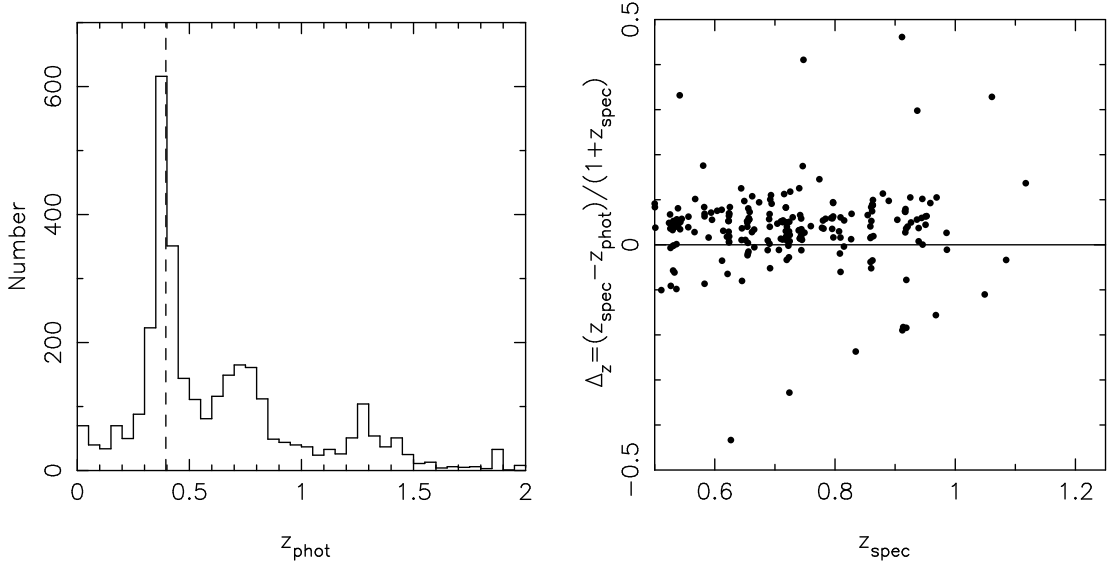

\centerline{
\psfig{file=f2a.ps,height=72mm,angle=0}
\hspace{5mm} \psfig{file=f2b.ps,width=70mm,angle=0} }

\caption{\small {\sc left:} Photometric redshift distribution in
the field of the cluster Cl\,0024 derived from the spectral
template fitting using {\sc hyperz}. The foreground cluster at
$z{=}0.4$ is clearly recovered, in addition to a significant
population of galaxies at $0.75{\le}z{\le}1.25$ which forms the
basis of this study. {\sc right:} Comparison of photometric and
spectroscopic redshifts from the catalog of Moran et al.\ (2004,
in prep.). The comparison indicates a mean redshift error of
${\langle}{\Delta}_z{\rangle}{=}0.04$ and rms scatter of
${\sigma}_z{=}0.1$. See \S\ref{losig} for further details.
\label{fig-photz} }
\end{figure*}

We now turn to the low density environments, as probed by the
wide--field observations of Cl\,0024. The WIRC $J$--band mosaic is of
key importance here since it provides a reasonable match to
rest--frame $V$--band at $0.75{\le}z{\le}1.25$. We analyze this data
with SExtractor (Berton \& Arnouts 1996) excluding all sources that
lie close to diffraction spikes around bright stars, adjacent to a
small number of remaining cosmetic defects on the final reduced mosaic
and within $10''$ of the edge of the field of view. Monte Carlo
simulations were used to determine the completeness limits of the
$J$--band catalog. Scaled artificial point--sources that match the
seeing were inserted at random positions into the $J$--band mosaic and
examined using the same SExtractor configuration as above.  The $80\%$
completeness limit (equivalent to a 5${\sigma}$ detection limit) was
determined to be $J(5{\sigma}){=}21.1$. We then performed aperture
photometry for all of the $J$--detected sources using a 2--arcsec
diameter aperture on the seeing matched $BVRIJK$--band frames.
Finally, we removed several hundred stars from the multi--color
catalog based on their profile shapes to yield a final catalog of 4376
sources.  Using {\sc hyperz}\footnote{Available at
http://webast.ast.obs-mip.fr/hyperz} (Bolzonella et al.\ 2000), we
then fitted synthetic spectral templates (Bruzual \& Charlot 1998) to
all 4376 galaxies in the $BVRIJK$ photometric catalog,
adopting a Calzetti et al.\ (2000) extinction law, and allowing dust
extinction in each galaxy to be $A_V{\le}1.2$.

The resulting photometric redshift distribution in
Fig.~\ref{fig-photz} shows that the foreground cluster, Cl\,0024
($z{=}0.4$), is well recovered in the photometric redshift
analysis. The photometric redshift reliability at higher redshifts can
be gauged by comparing with the extensive spectroscopic catalog of
Moran et al.\ (2004, in prep.\ -- see also Czoske et al.\ 2001; Treu
et al.\ 2003). The overlap between the photometric and spectroscopic
catalogs is limited beyond $z{=}1$ because the wavelength coverage of
the spectroscopic observations (${\lambda}{\ls}0.75{\rm {\mu}m}$ --
e.g.\ Treu et al.\ 2003) was designed to locate cluster members at
$z{\simeq}0.4$.  Nonetheless, in the region of overlap the mean
photometric redshift error is ${\langle}{\Delta}_z{\rangle}=0.04$,
where ${\Delta}_z{\equiv}(z_{\rm spec}-z_{\rm phot})/(1+z_{\rm
spec})$.  The rms scatter, defined as
${\sigma}_z^2{\equiv}(N-1)^{-1}{\Sigma}(({\Delta}_z-{\langle}{\Delta}_z{\rangle})/(1+z_{\rm
spec}))^2$, where $N$ is the number of galaxies, is also small:
${\sigma}_z{=}0.1$.

The final step in constructing the $z{\simeq}1$ field sample is to
select all galaxies within a suitable redshift range chosen to yield
an adequate--sized sample for the field of view. We adopted a range
$0.75{\le}z{\le}1.25$. Down to a luminosity $M_V{\le}-21.2$ (see
\S\ref{strategy}) the combined photometric/spectroscopic catalog
yields a sample of 843 galaxies.

Determining the optimal redshift bin, ${\delta}z$, for estimating
the galaxy density is a trade--off between two effects. To avoid
spurious associations ${\delta}z$ should ideally be as small as
possible. However, given the use of photometric redshifts, it is
pointless making the bin smaller than the typical error in
estimated redshift. After some experimenting, at each field galaxy
position, the ten nearest neighbors within a redshift slice
(${\delta}z{=}{\pm}0.1$) centered on the best--fit photometric
redshift (or spectroscopic redshift where available) were located.
The corresponding area was then computed as described above
(\S\ref{hisig}). Two corrections were subsequently applied. First,
a field correction in each redshift slice was computed by scaling
the number of galaxies within the entire field--of--view in each
slice. This leads to a reduction in the value of $\Sigma$ at each
location. The second correction takes account of uncertainties in
the photometric redshifts. For simplicity we assume that these
uncertainties are normally distributed. Since the measured scatter
(${\sigma}_z(1{+}z){\simeq}0.2$) is somewhat larger than the width
${\delta}z{=}{\pm}0.1$ of the interval employed for the density
measurement, the local density measurements are underestimated by
a factor of ${\sim}2$. The morphology--density relation is very
flat at the densities probed by these data (Fig.~\ref{fig-tsig}),
therefore this correction for photometric redshift uncertainties
has a negligible effect on the final results.

\subsection{Morphological Classification}\label{morph}

The total number of $z{\simeq}1$ galaxies for which detailed
morphological information is available is 1257. This comprises all
957 members of the high--density cluster catalog (\S\ref{hisig})
and 300 members out of the total of 843 galaxies in the
low--density field catalog (\S\ref{losig}) which lie on the
sparse--sampled \emph{HST} mosaic of Cl\,0024 (\S\ref{space}).

Postage stamp images ($5''{\times}5''$) of all 1257 galaxies were
extracted and classification was performed using a scheme
comprising stellar/compact, early--type (E/S0), late--type (Sa and
later) and faint categories, patterned after that employed by Treu
et al.\ (2003) but with broader classes designed to take account
of the lower signal--to--noise ratio of the most distant galaxies
targeted by this study.  One of us (GPS) classified all 1257
galaxies, and a control sample comprising a sub--set of roughly
one third of the total sample was cross--classified by three of the
authors (GPS, TT, RSE).  The majority of the differences in the
latter test arose from difficulties in classifying unambiguously
bulge dominated galaxies as either E/S0 (i.e.\ early--types) or Sa
(i.e.\ late--types in our scheme).  We use these three independent
morphological catalogs to estimate the uncertainty in the early
type fraction (6\%) and add this in quadrature to the statistical
errors when presenting our final results in \S\ref{tsigma}.

\section{Results}\label{results}

\begin{figure*}
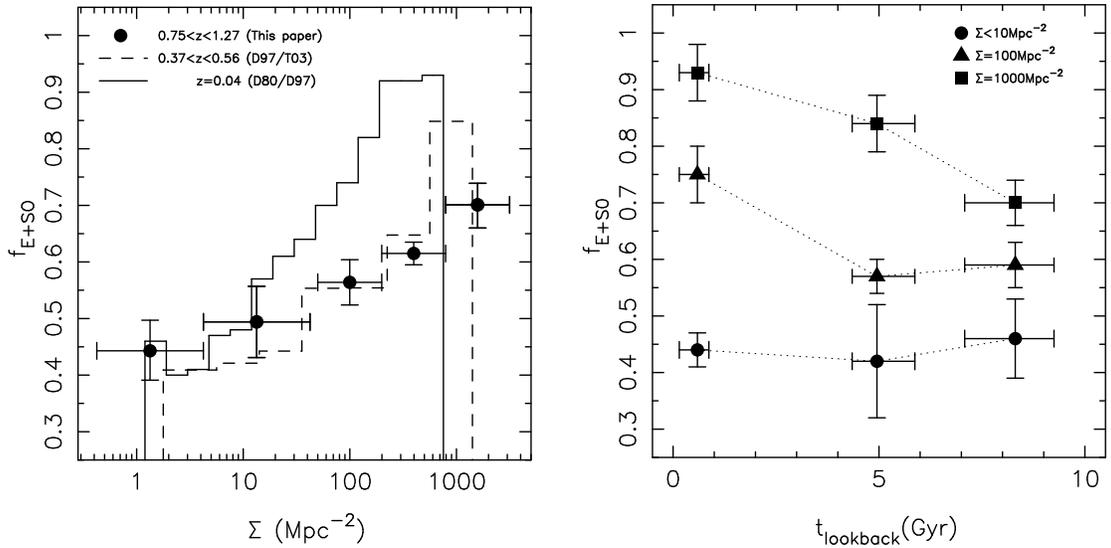

\centerline{ \psfig{file=f3a.ps,width=70mm,angle=0}
\hspace{5mm} \psfig{file=f3b.ps,width=70mm,angle=0}
}

\caption{\small {\sc left:} Early type fraction ${\feso}$ versus
projected density at various redshifts. Vertical error bars on the
filled data points represent the sum of uncertainties arises from
counting statistics, morphological misclassification, photometric
redshifts (for the two lowest density points) and field--to--field
variance (for the three high--density points). Horizontal error
bars define bin widths chosen to contain $>$100 galaxies. The
histograms show the low and intermediate redshift data presented
by Dressler et al.\ (1997); the $z{=}0.5$ data is rebinned to
include the results of Treu et al.\ (2003) and to achieve a
signal--to--noise ratio comparable with the high--redshift data.
{\sc right:} Evolution of the early--type fraction ${\feso}$ versus
look--back time for various projected densities derived from the
data presented on the left. \label{fig-tsig} }
\end{figure*}

\subsection{The Morphology--density Relation at $z{=}1$}\label{tsigma}

We now combine measurements of projected galaxy number density and
the morphological classifications to construct the
morphology--density relation at $z{=}1$ (Fig.~\ref{fig-tsig}). For
simplicity we summarize this relation in terms of the early--type
fraction, ${\feso}$, as a function of redshift and environmental
density.

Our data span three orders of magnitude in projected density from
the "field", ${\Sigma}{<}10\,{\rm Mpc^{-2}}$, to cluster cores,
${\Sigma}{\simeq}1000\,{\rm Mpc^{-2}}$.  The three highest density
points are derived from the pointed cluster observations
(\S\ref{hisig}); the two lowest density points are derived from
our analysis of the field viewed in the Cl\,0024 mosaiced image
(\S\ref{losig}). Horizontal error bars show bin widths chosen to
contains a minimum of $100$ galaxies (the ${\Sigma}{=}400\,{\rm
Mpc^{-2}}$ bin contains in excess of 600 galaxies).  Vertical
error bars combine binomial uncertainties (Gehrels et al.\ 1986)
with two further contributions added in quadrature.  First, we
quantify the cluster--to--cluster scatter by recomputing the high
density points, each time excluding one of the clusters (see
Table~\ref{obs}).  The rms scatter between these measurements of
${\feso}$ is ${\sim}0.03$, i.e.\ comparable with or smaller than
the typical binomial uncertainty.  We also include the effect of
morphological misclassifications as noted in \S\ref{morph}.

Fig.~\ref{fig-tsig} clearly shows that morphological segregation
was already present at $z{=}1$. The early type fraction ${\feso}$
monotonically increases with projected density, ${\Sigma}$.
Previous studies of the fraction of early--type galaxies at early
times concentrated on individual galaxy clusters (e.g.\ van Dokkum
et al.\ 2000, 2001; Lubin et al.\ 2002). These authors found
fractions consistent with those presented here when allowance is
made for the fact that averages were taken over larger areas (the
entire WFPC2 field in most cases) thereby sampling a range of
projected densities. Taking MS\,1054${-}$0321 as an example, van
Dokkum et al.\ (2000) found ${\feso}{=}0.5{\pm}0.1$ at densities of
${\Sigma}{\simeq}50$, which is consistent with the results
presented in Fig.~\ref{fig-tsig}.

\subsection{Evolution of the Morphology--density Relation}
\label{evolution}

Fig.~\ref{fig-tsig} also presents histograms of ${\feso}$ as a
function of projected density for the local and intermediate
samples, $z{\simeq}0$ and $z{\simeq}0.5$. The former is based on
Dressler et al.'s (1997) re--analysis of Dressler's (1980) data.
The latter combines Dressler et al.'s (1997) study of the core
regions of 10 clusters at $0.37{\le}z{\le}0.56$ with Treu et al.'s
(2003) panoramic study of Cl\,0024 ($z{=}0.4$). In combining these
two datasets, we re--binned Dressler et al.'s data to be
consistent with the Treu et al.\ data and took a simple average in
the region where the two datasets overlap, i.e.\
${\Sigma}{\ge}30\,{\rm Mpc^{-2}}$.

Although we detect a morphology--density relation at $z{=}1$, it
is not as prominent as in the local universe.  We quantify this
evolution by fitting a straight--line of the form
${\feso}{\propto}{\beta}\log{\Sigma}$ to both the $z{=}0$ and
$z{=}1$ data.  We obtain ${\beta}(z{=}0){=}0.26{\pm}0.01$ and
${\beta}(z{=}1){=}0.08{\pm}0.02$. The morphology--density
relation, as summarized by the early--type fraction, is therefore
${\sim}3$ times steeper locally than at $z{=}1$.

We also compare our $z{\simeq}1$ results with those at $z{=}0.5$,
and find, perhaps surprisingly, that there has been little
evolution between $z{=}1$ and $z{=}0.5$, except in the densest
bin, i.e.\ ${\Sigma}{\simeq}1000\,{\rm Mpc^{-2}}$. Fitting our
simple model to the $z{=}0.5$ data, we obtain
${\beta}(z{=}0.5){=}0.15{\pm}0.05$.  If we exclude the highest
density bin, the result changes only slightly:
${\beta}(z{=}0.5){=}0.13{\pm}0.05$.  Both of these values agree
within the uncertainties with the slope found at $z{=}1$.

A simpler way to present our results is the run of ${\feso}$ as a
function of look--back time for low (${\Sigma}{\le}10\,{\rm
Mpc^{-2}}$), intermediate (${\Sigma}{=}100\,{\rm Mpc^{-2}}$), and high
(${\Sigma}{=}1000\,{\rm Mpc^{-2}}$) densities (see
Fig.~\ref{fig-tsig}). This elucidates more clearly the timing of
environmental evolution.  Little evolution is seen in the early--type
fraction in low density environments over $0{<}z{<}1$.  Evolution at
intermediate densities occurred remarkably recently (i.e.\ in the last
5\,Gyr) with little evidence for any change at earlier times. In the
highest density regions, there has been a monotonic rise with cosmic
times.

\section{Discussion}\label{discuss}

We first consider why the fraction of early--type galaxies increases
first in the higher density environments, then in intermediate density
environments and finally -- if it does at all -- in the lowest density
environments.  Qualitatively, this can be understood in the scenario
of hierarchical structure formation.  At a given epoch, the densest
regions are those which started collapsing earliest; in terms of age
since collapse, the densest regions are therefore the oldest.  If we
assume that the original morphological mix is universal and then
late--type galaxies are transformed into early--types by environmental
processes, then the densest regions have had more time to increase
their early--type fraction. Clearly, the rate of transformation could
also be a function of density, for example dense clusters are likely
to be more efficient than poor groups at inducing ram--pressure
stripping, and therefore the metamorphosis could be accelerated once
some threshold conditions are met.  In summary, the broad picture
presented by our results is in qualitative agreement with the
hierarchical paradigm.  We now turn to more quantitative possible
explanations for our results.  We begin with a brief review of the
evolution of early--type galaxies in clusters.

For some years now, evidence spanning the range $0{<}z{<}1$ has
suggested that cluster early--types represent a very homogeneous,
slowly evolving population. This is based in part on the low intrinsic
scatter (${\sim}0.08$\,mags) observed in the local color--magnitude
relation (Bower, Lucey \& Ellis 1992) and that tracked to $z{\simeq}1$
(Ellis et al.\ 1997; Stanford, Eisenhardt \& Dickinson 1998).  The
mass--to--light ratios deduced from the fundamental plane provide a
second indicator, both at low redshift (e.g., Lucey et al. 1991;
Pahre, Djorgovski, \& de~Carvalho 1998) and intermediate redshifts
(e.g.\ van Dokkum \& Franx 1996; Bender et al. 1998; van Dokkum et
al. 1998; Kelson et al.\ 2000). Both results have supported the
widely--held view that the stars in {\it some} cluster early--types
formed at high redshift (i.e.\ $z{>}2$).

This does not necessarily mean that {\it all} local early--types
evolved from those seen at earlier times. Conceivably some formed
subsequent to $z{\simeq}0.5$--1 but nonetheless found their way onto
the present--day fundamental plane and color--magnitude relations
(Bower, Terlevich \& Kodama 1998). This is particularly likely for the
lenticulars which may have been transformed relatively recently from
star--forming galaxies (Dressler et al.\ 1997). However, the physical
processes that govern how star--forming disk galaxies are transformed
into quiescent lenticulars remains an important outstanding question
(e.g.\ Kodama \& Smail 2001; Treu et al.\ 2003).

Motivated by our new results, we now explore what new clues we can
deduce about the evolution of cluster early--type galaxies.
Specifically, we use several evolutionary scenarios to attempt to set
a limit on the fraction of lenticular galaxies, $\fso$, in clusters at
$z{=}1$.  Note that we restrict our attention to the high density
regions; this is because measurements of $\fso$ are not available at
lower redshift for the intermediate and low density regimes.  The crux
of our model is to use our measurement of $\feso$ at $z{=}1$, in
combination with the elliptical galaxy fraction, $\fe$, at $z{=}0.5$
(Dressler et al.\ 1997; Treu et al.\ 2003; \S\ref{evolution}) and
simple model assumptions to estimate $\fso$ at $z{=}1$.  We write the
following expression for the lenticular fraction at $z{=}1$:

\begin{equation}
f_{\rm S0,z=1}{=}f_{\rm E+S0,z=1}-f_{\rm E,z=0.5}\frac{N_{\rm z=0.5}}{N_{\rm z=1}}+\frac{{\Delta}N_{\rm E}}{N_{\rm z=1}}
\label{eq:frac}
\end{equation}

\noindent
We derive Equation~\ref{eq:frac} from first principles in the
Appendix, however it is quite straightforward to understand each term.
From the early--type fraction at $z{=}1$ ($f_{\rm E+S0,z=1}$), we
subtract the elliptical fraction at $z{=}0.5$, re--normalized to
account for changes in the total number of galaxies due to
evolutionary processes such as in--fall and galaxy--galaxy mergers.
We also add a term to account for changes in the number of elliptical
galaxies due to these evolutionary processes; we divide the change in
the number of ellipticals (${\Delta}N_{\rm E}{=}N_{\rm
E,z=0.5}{-}N_{\rm E,z=1}$) by the total number of galaxies at $z{=}1$.

We now employ a series of evolutionary scenarios from which we
estimate values of $N_{\rm z=0.5}/N_{\rm z=1}$ and ${\Delta}N_{\rm
E}/N_{\rm z=1}$, and thus, in combination with measurements of $f_{\rm
E+S0,z=1}$ and $f_{\rm E,z=0.5}$ derive estimates of $f_{\rm
  S0,z=1}$.  The numerical details of each scenario are listed in the
Appendix.  

We first adopt a closed box model in which we assume that
all cluster ellipticals are formed at high redshifts, say $z{>}2$, and
that the rising fraction of early--type galaxies (i.e.\ ellipticals
and lenticulars) with cosmic time arises entirely as a result of
lenticulars transformed from star--forming spirals.  A key prediction
of this model, and indeed the open box models discussed below, is the
existence of an epoch at which the early--type galaxy population in
clusters is ``pristine'', i.e.\ comprises solely ellipticals formed at
high redshift. Any measure of the fraction of lenticular galaxies
(${\fso}$) as a function of redshift would then yield important
constraints on the timing and the physics of galaxy transformation in
clusters.

For the closed box model, at $z{\ls}1$, ellipticals are neither
created nor destroyed (${\Delta}N_{\rm E}{=}0$) and there is no
overall number evolution ($N_{\rm z=0.5}{=}N_{\rm z=1}$).  The
lenticular fraction at $z{=}1$ is therefore simply the difference
between the early--type fraction at $z{=}1$ ($f_{\rm
E+S0,z=1}{=}0.7{\pm}0.1$) and the elliptical fraction at $z{=}0.5$
($f_{\rm E,z=0.5}{=}0.6{\pm}0.1$ -- Dressler et al.\ 1997; Treu et
al.\ 2003).  We therefore derive a crude upper limit of $f_{\rm
S0,z=1}{\ls}0.1$.  Given the uncertainties in the observational data,
in this picture, we could be witnessing such a ``pristine'' population
of cluster ellipticals at $z{\simeq}1$.  However, clusters are
probably not closed boxes; numerical simulations demonstrate that
material is continually accreted into clusters, generally along the
filamentary structure.  We therefore also explore several open box
models, with the aim of finding out whether additional evolutionary
processes tend to increase or decrease the closed box estimate of
$f_{\rm S0,z=1}$ .

First, we relax the assumption that there is no in--fall from the
field; we retain the assumption that there is no number evolution in
the ellipticals (${\Delta}N_{\rm E}{=}0$).  If we assume that the
$z{=}1$ cluster galaxy population has increased by 20\% at $z{=}0.5$
due to in--fall of spirals and lenticulars, then $N_{\rm z=0.5}/N_{\rm
z=1}{=}1.2$ and $f_{\rm S0,z=1}{\simeq}0$.  Note that this scenario
includes implicitly the possibility that the in--falling spirals are
transformed into lenticulars.  This simple in--fall scenario therefore
supports the idea that ${\fso}$ is negligible at $z{=}1$.

We now consider number evolution in the elliptical galaxies; his could
occur through several processes, for example some of the in--falling
population could already be ellipticals, spiral and/or lenticulars
could merge to form ellipticals either in the cluster core or in the
in--falling groups (e.g.\ van~Dokkum et al.\ 1999) and ellipticals in
the cluster cores could merge together to form a brightest cluster
galaxy (hereafter BCG; e.g.\ Nipoti et al.\ 2003).  Taking the
possibility of in--falling ellipticals first, we add 10\% in--fall of
ellipticals to the 20\% in--fall of spirals and lenticulars described
above: ${\Delta}N_{\rm E}/N_{\rm z=1}{=}0.1$; $N_{\rm z=0.5}/N_{\rm
z=1}{=}1.3$.  Substituting these values into Equation~\ref{eq:frac}
reveals that this scenario is also consistent with a very low
lenticular fraction at $z{=}1$ -- $f_{\rm S0,z=1}{\simeq}0.02$.

We now include galaxy--galaxy mergers as a mechanism for generating
cluster ellipticals, and for simplicity assume zero in--fall from the
field.  If ten in every hundred cluster spirals at $z{=}1$ merge
pair--wise to produce half that number of ellipticals by $z{=}0.5$,
then $N_{\rm z=0.5}/N_{\rm z=1}{=}0.95$ and ${\Delta}N_{\rm E}/N_{\rm
z=1}{=}0.05$, which translates into $f_{\rm S0,z=1}{\simeq}0.2$.
Combining this scenario with in--fall of a similar fraction of spiral
galaxies to that discussed above modifies the second term in
Equation~\ref{eq:frac} thus: $N_{\rm z=0.5}/N_{\rm z=1}{\simeq}1.2$,
and the lenticular fraction thus: $f_{\rm S0,z=1}{\simeq}0.03$.

Finally, under a galactic cannibalism scenario (e.g.\ Nipoti et al.\
2003), the number of cluster ellipticals reduces with time due to
their ingestion into the BCG.  If 5 per cent of cluster ellipticals at
$z{=}1$ have been cannibalized by $z{=}0.5$, then $N_{\rm
  z=0.5}/N_{\rm z=1}{\simeq}0.97$ and ${\Delta}N_{\rm E}/N_{\rm
  z=1}{=}{-}0.03$, which translates into $f_{\rm S0,z=1}{\simeq}0.09$.
Again, adding 20 per cent in--fall of spiral galaxies to a cannibalism
scenario yields $N_{\rm z=0.5}/N_{\rm z=1}{\simeq}1.17$, and a
lenticular fraction of: $f_{\rm S0,z=1}{\simeq}0$.  The demonstrates
that it is unreasonable to assume that the agent of change is only the
spiral population and that a combination of cannibalism and in-fall in
the open box case can be arranged to yield a low lenticular fraction
at $z{=}1$.

In summary, we have used simple models to explore several scenarios
for the evolution of early--type galaxies between $z{=}1$ and
$z{=}0.5$, with the aim of constraining the fraction of lenticular
galaxies in clusters at $z{=}1$.  Whilst the scenarios considered are
unlikely to represent an exhaustive study, it is interesting to note
that in all except one scenario the lenticular fraction is
${\fso}{=}0.1$ or lower.  This is comparable with the uncertainty on
the observational data included in the calculations using
Equation~\ref{eq:frac}.  At $z{=}1$, we may therefore be observing
cluster galaxy populations at or very close to their ``pristine''
state, in a scenario where the bulk of the elliptical population
formed at higher redshifts ($z{>}2$).

Our suggestion that the lenticular fraction at $z{=}1$ is negligible
is clearly speculative. Additional data is required to test this
interpretation, most importantly, a discriminator between elliptical
and lenticular galaxies at high redshift is required. In addition to
deep \emph{HST}/ACS imaging for morphologies, resolved spectroscopy of
early--type galaxies in clusters at $z{\simeq}1$ and beyond should
help to discriminate between those galaxies that are dynamically hot
(elliptical galaxies) and those that are cold, i.e.\ lenticular
galaxies with systematic rotation. Already, promising exploratory
studies have demonstrated the feasibility of making this distinction
(van Dokkum \& Stanford 2001, Iye et al.\ 2003).

\section{Conclusions}\label{conc}

We have used 52 individual \emph{HST}/WFPC2 observations through the
F814W filter, supplemented by panoramic ground--based imaging to
measure the morphology--density relation of galaxies at $z{=}1$.  Our
study adopts analysis methods similar to those developed at lower
redshifts (e.g.\ Dressler 1980) and our principal achievement is to
span, at $z\simeq$1, the full three orders of magnitude range in the
projected number density of galaxies encompassed by the low redshifts
studies. We choose to make a like--for--like comparison of the
early--type fractions spanning field (${\Sigma}{\ls}$10\,Mpc$^{-2}$),
group (${\Sigma}{\simeq}$100\,Mpc$^{-2}$) and rich cluster
(${\Sigma}{\simeq}$1000\,Mpc$^{-2}$) environments.

We briefly summarize our findings as follows:

\begin{list}{(\roman{fred})}{\usecounter{fred}\setlength{\labelwidth}{5mm}\setlength{\itemindent}{0mm}\setlength{\labelsep}{1.5mm}\setlength{\leftmargin}{6.5mm}\setlength{\itemsep}{2mm}} 

\item Morphological segregation remains a prominent feature of the
galaxy population at $z{=}1$, although the slope of the
${\feso}$--${\log}{\Sigma}$ relation is ${\sim}3$ times shallower than 
observed locally.

\item The morphology--density relations at $z{=}1$
and $z{=}0.5$ are remarkably similar, with a significant
difference only detected in the highest density bin. Most of the
evolution producing the locally--observed relation occurred in the
redshift interval 0$<z<$0.5.

\item At low densities, the early--type fraction is
roughly constant at ${\feso}{=}0.4{\pm}0.1$ across the full
redshift range ($0{<}z{<}1$).

\end{list}

These trends suggest to us a simple model whereby most cluster
ellipticals formed at high redshift ($z{>}2$) with the bulk of the
density--dependent growth arising from the environmental transformation
of in--falling disk galaxies into lenticulars, and possibly merging of
cluster galaxies at later times. This is motivated by the suggestive
agreement (within the uncertainties) between the early--type fraction
at $z{=}1$ in high density regions with the elliptical fraction
observed at $z{=}0.5$.  Within the observational uncertainties, the
majority of the model scenarios that we have explored are consistent
with a negligible lenticular fraction at $z{=}1$, ${\fso}{\ls}0.1$.
It is therefore possible that all cluster early--types at $z{=}1$ are
ellipticals. To test this suggestion, resolved dynamical data is
needed for a large sample of early--type cluster and field galaxies
whose environmental densities can be measured.

\section*{Acknowledgments}

GPS thanks Andrew Benson, Pieter van~Dokkum and David Sand for helpful
discussions and comments.  We are grateful to Jean--Paul Kneib and
Oliver Czoske for generously sharing their ground--based optical data
with us.  We thank Kevin Bundy, David Thompson and John Carpenter for
helpful comments on the NIR data reduction and calibration, and thank
Chris Conselice for assistance with the NIR observations.  We also
recognize Ian Smail and David Gilbank's valiant efforts in 2000/2001
to obtain and reduce wide--field NIR imaging data of Cl\,0024.  We
acknowledge financial support for proposal HST--GO--8559.  TT also
acknowledges support from NASA through Hubble Fellowship grant
HF--01167.01.  Finally, we recognize and acknowledge the cultural role
and reverence that the summit of Mauna Kea has within the indigenous
Hawaiian community. We are most fortunate to have the opportunity to
conduct observations from this mountain.

\appendix

\section{Derivation of Equation~1}\label{deriv}

Equation~\ref{eq:frac} describes how the fraction of lenticular
galaxies at $z{=}1$ ($f_{\rm S0,z=1}$) can be estimated from two
observable quantities: the fraction of early--type galaxies at $z{=}1$
($f_{\rm E+S0,z=1}$) and the fraction of elliptical galaxies at
$z{=}0.5$ ($f_{\rm E+S0,z=0.5}$).  As explained in \S\ref{discuss},
the equation is quite intuitive, however for completeness, we derive
it here from first principals.  First, we write the early--type
fraction at $z{=}1$ in terms of the elliptical and lenticular
fractions at that redshift:

\begin{equation}
f_{\rm E+S0,z=1}{=}f_{\rm E,z=1}{+}f_{\rm S0,z=1}
\label{eq:2}
\end{equation}

\noindent Simple re--arrangement, where $N_{\rm z=1}$ is the total
number of cluster galaxies and $N_{\rm E,z=1}$ is the number of
cluster ellipticals, both at $z{=}1$, gives:

\begin{equation}
f_{\rm S0,z=1}{=}f_{\rm E+S0,z=1}{-}\frac{N_{\rm E,z=1}}{N_{\rm z=1}}
\label{eq:3}
\end{equation}

\noindent We now define ${\Delta}N_{\rm E}{=}N_{\rm E,z=0.5}{-}N_{\rm
 E,z=1}$ to be the change in the number of cluster ellipticals between
  $z{=}1$ and $z{=}0.5$, and re--write Equation~\ref{eq:3} as:

\begin{equation}
f_{\rm S0,z=1}{=}f_{\rm E+S0,z=1}{-}\frac{N_{\rm
    E,z=0.5}{-}{\Delta}N_{\rm E}}{N_{\rm z=1}} 
\label{eq:4}
\end{equation}

\noindent Finally, we substitute $N_{\rm E,z=0.5}{=}f_{\rm
E,z=0.5}.N_{z=0.5}$, to obtain Equation~\ref{eq:frac} from
\S\ref{discuss}:

\begin{equation}
f_{\rm S0,z=1}{=}f_{\rm E+S0,z=1}{-}f_{\rm E,z=0.5}\frac{N_{\rm z=0.5}}{N_{\rm z=1}}{+}\frac{{\Delta}N_{\rm E}}{N_{\rm z=1}}
\label{eq:5}
\end{equation}

\section{Tabulation of Closed and Open Box Models}\label{models}

In this table we list the values used in Equation~\ref{eq:frac} in \S\ref{discuss}.

\begin{table*}[h]
\label{tab:mods}
{\scriptsize
\begin{center}
\begin{tabular}{lccccccccc}
\hline \noalign{\smallskip} 
 & $f_{\rm S0,z=1}$ & = & $f_{\rm E+S0,z=1}$ & $-$ & $f_{\rm E,z=0.5}.N_{\rm z=0.5}/N_{\rm z=1}$ & $+$ & ${\Delta}N_{\rm E}/N_{\rm  z=1}$ & & \cr
\noalign{\smallskip} \hline
\noalign{\smallskip}
Closed Box Model \cr
~[A]~~No in--fall, no number evolution & $f_{\rm S0,z=1}$ & = & 0.7 & -- & $0.6{\times}1$ & + & 0 & = & 0.1 \cr
\noalign{\smallskip}
Open Box Models \cr
~[B]~~20\% in--fall of spirals and lenticulars & $f_{\rm S0,z=1}$ & = & 0.7 & -- & $0.6{\times}1.2$ & + & 0 & = & $-0.02$ \cr
~[C]~~Model B plus 10\% in--fall of ellipticals & $f_{\rm S0,z=1}$ & = & 0.7 & -- & $0.6{\times}1.3$ & + & 0.1 & = & $0.02$ \cr
~[D]~~10\% of the total population (assumed to be   & $f_{\rm S0,z=1}$ & = & 0.7 & -- & $0.6{\times}0.95$ & + & 0.05 & = & $0.18$ \cr
~~~~~~~spirals) merge pair--wise to form ellipticals\cr
~[E]~~Model D plus Model B & $f_{\rm S0,z=1}$ & = & 0.7 & -- & $0.6{\times}1.2$ & + & 0.05 & = & $0.03$ \cr
~[F]~~Cannibalism -- 5\% of ellipticals merge to   & $f_{\rm
 S0,z=1}$ & = & 0.7 & -- & $0.6{\times}0.97$ & $-$ & 0.03 & = & $0.09$
 \cr
~~~~~~~form a BCG\cr
~[G]~~Model F plus Model B & $f_{\rm S0,z=1}$ & = & 0.7 & -- & $0.6{\times}1.17$ & $-$ & 0.03 & = & $-0.03$ \cr
\noalign{\smallskip} \hline \noalign{\smallskip}
\end{tabular}
\end{center}
}
\end{table*}

\end{document}